\documentclass[preprint,aps]{revtex4}
\begin{document}
\title{DIRAC EQUATION WITH VECTOR AND SCALAR POTENTIALS VIA SUSY QM}
\author{E. S. Rodrigues$^{(a,b)},$ A. F. de Lima$^{(a)}$ and R. de Lima Rodrigues$^{\beta}$\\
{}$^{(a)}$Unidade Acad\^emica de F\'\i sica, Universidade Federal de
Campina Grande \\
58109-790 Campina Grande - PB, Brazil\\
{}$^{\beta}$Unidade Acad\^emica de Educa\c{c}\~ao, Universidade Federal de
Campina Grande\\
58.175-000 Cuit\'e - PB, Brazil\\
{}$^{(b)}$Departamento de F\'\i sica, Instituto Federal de Educa\c{c}\~ao, Ci\^encia e Tecnologia do Sert\~ao de Pernambuco - Campus Salgueiro\\
Cidade Universit\'aria, 56.000-000
Salgueiro - PE, Brazil}


\begin{abstract}
In this work, a spin $\frac 12$ relativistic particle described by a
generalized potential containing both the Coulomb potential and
a Lorentz scalar potential in Dirac equation is investigated in terms of
the generalized ladder operators from  supersymmetry in quantum mechanics.
 This formalism is
applied for the generalized Dirac-Coulomb problem which is an
exactly solvable potential in relativistic quantum mechanics. We obtain the
energy eigenvalues and calculate explicitly the energy eigenfunctions for
the ground state and first excited state.
 
 \vspace{1cm}


\vspace{3cm} e-mails: to
ESR is eriverton.rodrigues@ifsertao-pe.edu.br, to AFL is aerlima@df.ufcg.edu.br and to RLR are rafaelr@cbpf.br or
rafael@df.ufcg.edu.br. 
\end{abstract}

\maketitle

\newpage

\section{Introduction}

\paragraph*{}

Using first the separation in polar coordinates system for the generalized Dirac equation, in three dimensions, the supersymmetry (SUSY) in quantum mechanics (QM) is investigated. We consider a spin $\frac 12$ particle in the Coulomb potential, $V_c= -\frac{A_1}{r}$, and a Lorentz
scalar potential, added to the mass
term in the Dirac equation which may be interpreted as an effective
position dependent mass. If the scalar potential is assumed to be
created by the exchange of massless scalar mesons, it has the form
$V_s= -\frac{A_2}{r}.$

The (3+1) and (1+1) dimensional Dirac equations with both
scalar-like and vector-like potentials have been well known in the
literature for a long time \cite{Zhong85}. Exact solutions for the
bound states in this mixed potential can be obtained by the method
of separation of variables \cite{gre90,Tutik92,Ik93,bose01} and also
by the use of the dynamical algebra $SO(2,1)$ \cite{pan95}. In a
recent paper the solution  of the scattering problem for this
potential has been obtained by an analytic method and also by an
algebraic method by Vaidya and Silva Souza \cite{AL02}.

We do not think that the solution of the relativistic Coulomb potential
with position-dependent mass in the Coulomb field
has been correctly solved.

Recently exact solutions have been found for fermions in the presence
of a classical background which is a mixing of the time-dependent of
a gauge potential and a scalar potential \cite{chen05}. Also,
exactly solvable Eckart scalar and vector potentials in the Dirac
equation have been investigated via SUSY QM \cite{jia05}, the
$S$-wave Dirac equation has been solved exactly for a single
particle with spin and pseudospin symmetry moving in a central
Woods-Saxon potential \cite{guo05}, and it was shown that some other
cases for which the Dirac equation with classical potentials of
vector and scalar natures with spherical asymmetry can be solved
exactly \cite{alvaro07}.

In this work we consider an alternative calculation for the energy
spectrum and eigenfunctions of the Dirac equation via a
Schr\"odinger-like wave equation, for the vector and scalar
potentials recently studied by Alhaidari \cite{alha05}. The connection between the Johnson-Lippmann \cite{Stahlhofen97,Martinez05} and the generalized Johnson-Lippman operators \cite{Leviatan04,Tamar07} in a Dirac-Coulomb
problem via SUSY QM has been studied in 3-dimensional space. In \cite{Stahlhofen97}
relations between various algebraic approaches  via radial
equation are pointed out, and in \cite{Martinez05} the
bound eigenfunctions and spectra of a Dirac hydrogen atom have been
found via $su(1,1)$ Lie algebra.

 Our calculation generalizes the results previously
obtained for the relativistic Coulomb
problem. Instead of a direct generalization of Sukumar's calculation
we use the equivalent approach developed by Fukui and Aizawa \cite
{fukui93} and also by Balantekin \cite{Bala98,Armando01} where shape
invariance \cite {Gend83} plays an important role. The Coulomb
problem in non-relativistic quantum mechanics has been obtained by
the use of SUSY shape invariances \cite{Elso-01}. The special case
of the Dirac-Coulomb potential has been treated recently via SUSY
quantum mechanics by one of the authors \cite{R04}.

This work is organized as follows. In section II, from the time independent Dirac
equation for a  potential in terms of the Coulomb potential and
a Lorentz scalar potential we construct the radial equations. In section
III, we have achieved the diagonalization  of the matrix that appears in the interaction term of the radial Dirac equation, supercharges, supersymmetric
Hamiltonian and the  
ground state energy eigenvalue and eigenfunction.
In section IV, using the shape invariance properties  we deduce  generalised
ladder operators to build up
the bound state energy eigenvalues and eigenfunctions. The conclusion is
presented in section V.

\section{Time independent Dirac equation}

The time independent Dirac equation with vector $V(r)$ and scalar
potentials $V_S(r)$ may be written in the form

\begin{equation}
H\Psi= E\psi,
\end{equation}
where the Hamiltonian is given by

\begin{equation}
\label{h1} H= \rho_1\otimes\vec\sigma\cdot\vec p +\rho_3\otimes \bf
{1}_{2\hbox{x}2}(M+V_S(r)) +V(r) \bf
{1}_{4\hbox{x}4},
\end{equation}
and we have used a direct product notation in which $\rho_i$ and
$\sigma_i, (i= 1,2,3)$ are the Pauli spin matrices obeying $[\rho_i,
\sigma_j]_-= 0,$ with $\hbar=1=c.$

The Hamiltonian (\ref{h1}) commutes with the total angular 
momentum

\begin{equation}
\vec J= \vec L+\frac 12 \vec\sigma.
\end{equation}
It also commutes with the Dirac operator

\begin{equation}
K= \rho_3{\bf (1}+\vec\sigma\cdot \vec L),
\end{equation}
and with the inversion operator

\begin{equation}
P= \rho_3 I,
\end{equation}
where $I$ inverts the spatial coordinates and the momenta.

The symmetry group of this Hamiltonian (\ref{h1}) is characterized by the two 
vector invariants $J$ e $K$, precisely as in the nonrelativistic case.
A complete set of mutually commuting operators is $H, J^2, J_3, K,
P$ which have the simultaneous eigenvectors  $\mid E, j, m,
\kappa>$ with

\begin{eqnarray}
J^2\mid E, j, m, \kappa>&&= j(j+1)\mid E, j, m, \kappa>\nonumber\\
J_3\mid E, j, m, \kappa>&&= m\mid E, j, m, \kappa>\nonumber\\
K\mid E, j, m, \kappa>&&= -k\mid E, j, m, \kappa>
\end{eqnarray}
where $j$ half integral, $m= -j, \cdots j$ and $k= \pm\left(j+\frac
12\right).$

To separate variables we introduce two component spinors which are
eigenfunctions of  $J^2, J_z, L^2, S^2$ and are of two types

\begin{equation}
\phi^{(+)}_{j, m}= \left(\begin{array}{cc}\left(\frac{\ell +\frac
12 +m}
{2l+1}\right)^{\frac 12}Y_{\ell, m-\frac 12}\\
\left(\frac{\ell+\frac 12 -m}{2l+1}\right)^{\frac 12}Y_{\ell,
m+\frac 12}
\end{array}\right),
\end{equation}
for $j= \ell+\frac 12, k=j+\frac 12>0$ and

\begin{equation}
\phi^{(-)}_{j, m}= \left(\begin{array}{cc}\left(\frac{\ell+\frac
12 -m}
{2l+1}\right)^{\frac 12}Y_{\ell, m-\frac 12}\\
-\left(\frac{\ell+\frac 12 +m}{2l+1}\right)^{\frac 12}Y_{\ell,
m+\frac 12}
\end{array}\right),
\end{equation}
for $j= \ell-\frac 12, k=-(j+\frac 12).$

In the above basis one verify that

\begin{equation}
J^2\phi^{(\pm)}_{jm}= j(j+1)\phi^{(\pm)}_{jm},
\end{equation}

\begin{equation}
\label{n1}
\vec\sigma\cdot\vec n\phi^{(\pm)}_{jm}= \phi^{(\pm)}_{jm},
\end{equation}
where $\vec n= \frac{\vec r}{r},$ and

\begin{equation}
{\bf (1}+\vec\sigma\cdot \vec L)\phi^{(\pm)}_{jm}=
-k\phi^{(\pm)}_{jm},
\end{equation}
where $k= \pm\left(j+\frac 12\right)$ for $j= \ell\mp\frac 12.$

Next, we put

\begin{equation}
\Psi= \left(\begin{array}{cc}
\frac{iG_{\ell j}}{r}\phi^{\ell}_{jm}\\
\frac{F_{\ell j}}{r} \vec\sigma\cdot\vec  n\phi^{\ell}_{jm}
\end{array}\right),
\end{equation}
where

\begin{equation}
\phi^{\ell}_{jm}=\phi^{(\pm)}_{jm},
\end{equation}
for $j= \ell\pm\frac 12.$ The operator $\vec\sigma\cdot\vec  n$ when expressed in this representation, as was done in (\ref{n1}), has the property of
reversing the sign of $\kappa$. The result which can be easily established by means of this operator relation

\begin{equation}
[{\bf 1}+\vec\sigma\cdot\vec L, \vec\sigma\cdot\vec n]_+= 0.
\end{equation}
Next, using this relation we have

\begin{equation}
K\Psi= -k\Psi.
\end{equation}

Further, using the relations

\begin{equation}
\vec\sigma\cdot \vec p\frac{f(r)}{r}\phi^{\ell}_{jm}=
-\frac{i}{r}\left(\frac{df}{dr}+\frac{kf}{r}\right)\vec\sigma\cdot\vec
n \phi^{\ell}_{jm},
\end{equation}
and

\begin{equation}
\vec\sigma\cdot\vec p\vec\sigma\cdot\vec
n\frac{f(r)}{r}\phi^{\ell}_{jm}=
-\frac{i}{r}\left(\frac{df}{dr}-\frac{kf}{r}\right)
\phi^{\ell}_{jm},
\end{equation}
we get the radial equations

\begin{eqnarray}
\frac{dG_{\ell j}}{dr}+\frac{k}{r}G_{\ell j}-\left(E+M+V_S(r)+
V(r)\right)F_{\ell j}&&= 0,\nonumber\\
\frac{dF_{\ell j}}{dr}-\frac{k}{r}F_{\ell j}+\left(E-M-V_S(r)+
V(r)\right)G_{\ell j}&&= 0,
\end{eqnarray}
where $k$ is the eigenvalue of the Dirac operator $K.$

Defining

\begin{eqnarray}
\Phi= \left(\begin{array}{cc}G_{\ell j}\\F_{\ell
j}\end{array}\right),
\nonumber\\
\Lambda=\pm \kappa\rho_3 -rV_S(r)\rho_1 - irV(r)\rho_2,\nonumber\\
 \vec k\cdot\vec\rho= M\rho_1 + iE\rho_2,
\end{eqnarray}
the Dirac radial equation for the hydrogen atom may be written in the form

\begin{equation}
\left[\frac{d}{dr}+\frac{\Lambda}{r}-\vec
k\cdot\vec\rho\right]\Phi= 0.
\end{equation}

Next we show that as the Dirac equation becomes a Schr\"odinger-like
wave equation, with the vector and scalar potentials.

\section{Energy spectrum and eigenfunctions via SUSY}

Now, let $S$ be the operator which diagonalises the matrix that appears in the interaction term. The matrix $\Lambda$ may be diagonalized by using the result

\begin{equation}
S^{-1}\Lambda S= \lambda\rho_3,
\end{equation}
where
\begin{equation}
\lambda^2=\kappa^2-A_1^2+A_2^2
\end{equation}
and

\begin{equation}
S= \left(\begin{array}{cc}c & d\frac{A_1 - A_2}{\lambda +
\mid\kappa\mid}\\ c\frac{A_1 + A_2}{\lambda +
\mid\kappa\mid}&d\end{array} \right).
\end{equation}

Defining

\begin{equation}
\Phi= S\hat{\Phi},
\end{equation}
we get

\begin{equation}
\label{Ek2} \left[\frac{d}{dr}+\frac{\lambda}{r}\rho_3 - \vec{\hat
k}\cdot\vec\rho\right]\hat{\Phi}= 0,
\end{equation}
where

\begin{eqnarray}
2\frac cd\lambda(\mid\kappa\mid +\lambda)\hat{k}_- &&=
(\mid\kappa\mid +\lambda)^2(M+E)-(A_1 - A_2)^2(M-E)\nonumber\\
2\frac dc\lambda(\mid\kappa\mid +\lambda)\hat{k}_+ &&=
(\mid\kappa\mid +\lambda)^2(M-E)-(A_1 + A_2)^2(M+E)\nonumber\\
\hat{k_3}&&= \frac{EA_1 + MA_2}{\lambda}.
\end{eqnarray}

Thus, we get

\begin{eqnarray}
\label{k}
-\hat{k}_- \hat{k}_+ &&= \frac{a^2}{\lambda^2}\nonumber\\
{\hat{k}}^2M^2 &&= k^{2}= M^2 - E^2
\end{eqnarray}
where

\begin{equation}
\label{a}
M^2a^2=(EA_1+MA_2)^2-k^2\lambda^2,
\end{equation}
which provides a relation between the SUSY and an additional constant of
motion in the Dirac-Coulomb problem will be reported separately.

It may be noted that Eq. (\ref{Ek2}) can be rewritten as

\begin{equation}
\label{A+-} A^+\hat{G}= \hat{k}_-\hat{F}, \quad A^-\hat{F}=
-\hat{k}_+\hat{G}
\end{equation}
where

\begin{equation}
\label{AA+-}
A^{\pm}= \pm\frac{d}{dr}+\frac{\lambda}{r}-\frac{EA_1 +
MA_2}{\lambda}.
\end{equation}
These relations are similar to the relations between the two
components of the eigenfunctions of a supersymmetric Hamiltonian
\cite{witten81}

\begin{eqnarray}
{\cal H}&&= [{\bf Q, Q^{\dagger}}]_+={\bf QQ^{\dagger}}+{\bf Q^{\dagger}Q}\nonumber\\
{}&&= \left(\begin{array}{cc} A^+ A^- & 0\\
0 & A^- A^+\end{array}\right)= \left(\begin{array}{cc} {\cal H}_- & 0\\
0 & {\cal H}_+\end{array}\right)\nonumber\\
{\bf Q^{\dagger}}&&= \left(\begin{array}{cc}0 & A^+\\0 &
0\end{array}\right), \quad [{\bf Q}, {\cal H}]_-=0= [{\bf
Q^{\dagger}},{\cal H}]_-,
\end{eqnarray}
where $\bf Q$ and $\bf Q^{\dagger}$ are the supercharges satisfying the anticommutation
relations of the SUSY algebra.
The supersymmetric partner Hamiltonians ${\cal H}_{\pm}$ satisfy the
following eigenvalue equations: ${\cal
H}_{\pm}\psi^{(n)}_{\pm}=E^{(n)}_{\pm}\psi^{(n)}_{\pm}.$ Next, Eq.
(\ref{A+-}) may be written as

\begin{eqnarray}
\label{EA+A-}
A^- A^+\hat{G}= \frac{a^2}{\lambda^2}M^2\hat{G}, \quad
A^+ A^-\hat{F}= \frac{a^2}{\lambda^2}M^2\hat{F}
\end{eqnarray}
which  shows  that every eigenvalue of $A^+ A^-$ is also an
eigenvalue of $A^- A^+$ except when $A^-\hat{F}= 0.$ This
condition gives the lower component for the "ground state" wave function

\begin{equation}
\label{F0}
\hat{F}_{0}= r^{\lambda}e^{-\frac{(E_0 A_1 +
MA_2)r}{\lambda}}
\end{equation}
and 

\begin{eqnarray}
{E_0\over M}=\pm\left( 1-{{A_1^2+A_2^2}\over \lambda^2+A_1^2}+
{{A_1^2A_2^2}\over  \left(\lambda^2+A_1^2\right)^2}\right)^{1\over
2} -{{A_1A_2}\over \lambda^2+A_1^2},
\end{eqnarray}
where $E_0$ is the ground state energy eigenvalue, for $a^2=0$.

\section{Excited states via Shape invariance condition}

The shape-invariant SUSY partner potentials
\cite{Gend83,CGK,Sukhat88,Fred}
are similar in shape and differ only in the
parameters that appear in them. More specifically, if $V_{-}(x;a_{1})$ is
any potential, adjusted to have zero ground state energy $E^{(0)}_{-} = 0,$
its SUSY partner
$V_{+}(x;a_{1})$ must satisfy the requirement

\begin{equation}
V_+ (x;a_1) = V_- (x;a_2) + R(a_2), \qquad a_2 = f(a_1),
\label{E56}
\end{equation}
 
where $a_{1}$ is a set of parameters, $a_{2}$ a function of the
parameters $a_{1}$ and $R(a_{2})$ is a remainder independent of $x$.
Then, starting with $V_{1} = V_{-}(x;a_{2})$ and
$V_{2} = V_{+}(x;a_{1}) = V_{1}(x;a_{2}) + R(a_{2})$ in (\ref{E56}),
one constructs a hierarchy of Hamiltonians

\begin{equation}
H_n = -\frac{1}{2}\frac{d^{2}}{dx^{2}} + V_{-} (x;a_n) +
\Sigma_{s=2}^{n} R(a_s),
\label{E57}
\end{equation}
where $a_{s} = f^{s}(a_{1})$, i.e., the function $f$
applied $s$ times. In view of Eqs. (\ref{E56}) and (\ref{E57}), we have

\begin{equation}
H_{n+1}
= -\frac{1}{2}\frac{d^{2}}{dx^{2}} + V_{-}(x;a_{n+1}) +
\Sigma_{s=2}^{n+1}R(a_s)
\label{E58a}
\end{equation}

\begin{equation}
= -\frac{1}{2}\frac{d^{2}}{dx^{2}} + V_{+}(x;a_{n}) + \Sigma_{s=2}^{n}R(a_s).
\label{E58b}
\end{equation}
Comparing (\ref{E57}), (\ref{E58a}) and (\ref{E58b}), we immediately note
that $H_{n}$ and
$H_{n+1}$ are SUSY partner Hamiltonians with identical energy spectra
except for the ground state level

\begin{equation}
E^{(0)}_n = \Sigma_{s=2}^{n}R(a_s)
\label{E59}
\end{equation}
of $H_{n}$, which follows from Eq. (\ref{E57}) and the
normalization that for any
$V_{-}(x;a)$ , $E^{(0)}_{-} = 0$. Thus, we get

\begin{equation}
E^{n}_1 = E^{n-1}_{2} = \ldots = E^{(0)}_{n+1} = \sum_{s=2}^{n+1}R(a_s),
\quad
 n = 1, 2, \ldots
\label{E60a}
\end{equation}
and

\begin{equation}
\psi^{(n)}_{1}\propto A^{+}_1(x;a_1) A^{+}_2(x;a_2) \ldots A^{+}_n(x;a_n)
\psi^{(0)}_{n+1}(x;a_{n+1}).
\label{E60b}
\end{equation}

Equations (\ref{E60a}) and (\ref{E60b}), succinctly express the SUSY
algebraic generalization, for various shape-invariant potentials of
physical interest \cite{Gend83,Sukhat88}, of the method of
constructing energy eigenfunctions $(\psi^{(n)}_{osc})$ for the usual ID
oscillator problem. Indeed, when $a_1=a_2= \ldots =a_n=a_{n+1},$ we obtain
$\psi^{(n)}_{osc}\propto (a^+)^n \psi^{(0)}_{1}, \quad A_1^+=\cdots A^+_n=a^+, \quad
\psi^{(0)}_{osc}=\psi^{(0)}_{n+1}=\psi^{(0)}_1 \propto e^{-\frac{\omega
x^2}{2}},$ where $\omega$ is the angular frequency.

The shape invariance has an underlying algebraic structure and may
be associated with Lie algebra \cite{Armando01}.
Now, we present our own application of the shape invariant method outlined above. Consider next the supersymmetric partner Hamiltonians given by

\begin{eqnarray}
{\cal H_-}&=& A^+ A^- =
-\frac{d^2}{dr^2}+\frac{\lambda(\lambda-1)}{r^2}-\frac{2(EA_1+MA_2)}{r}
+\frac{(EA_1+MA_2)^2}{\lambda^2}\nonumber\\
&\equiv& -\frac{d^2}{dr^2}+V_-(r,\lambda)
\end{eqnarray}
and 

\begin{eqnarray}
{\cal H_+}&=& A^- A^+=
-\frac{d^2}{dr^2}+\frac{\lambda(\lambda+1)}{r^2}
-\frac{2(EA_1+MA_2)}{r}+\frac{(EA_1+MA_2)^2}{\lambda^2}\nonumber\\
&\equiv& -\frac{d^2}{dr^2}+V_+(r,\lambda).
\end{eqnarray}
These SUSY partner potentials are shape invariant, since

\begin{equation}
\label{V22}
V_+(r,\lambda)= V_-(r,\lambda+1)+R(\lambda+1),
\end{equation}
where

\begin{equation}
\label{R}
R(\lambda+1)=\frac{(EA_1+MA_2)^2}{\lambda^2}
-\frac{(EA_1+MA_2)^2}{(\lambda+1)^2}.
\end{equation}
Hence

\begin{equation}
\label{sia}
A^-(\lambda)A^+(\lambda)=A^+(\lambda+1)A^-(\lambda+1)+
R(\lambda+1).
\end{equation}

Following the approach of Fukui and Aizawa \cite{fukui93} and also
of Balantekin \cite{Bala98,Armando01}, we define the following ladder
operators

\begin{equation}
\label{B-}
B^-(\lambda)= T^{\dagger}(\lambda)A^-(\lambda), \quad
B^+(\lambda)= (B^-)^{\dagger}(\lambda),
\end{equation}
where $T(\lambda)$ is a translation operator defined by

\begin{equation}
\label{T}
T(\lambda)= e^{\frac{\partial}{\partial \lambda}},
\end{equation}
with  $T^{\dagger}(\lambda)= e^{-\frac{\partial}{\partial
\lambda}},$ and get

\begin{equation}
\lbrack B^-(\lambda),B^+(\lambda) \rbrack = R(\lambda).
\end{equation}
It is easy to verify that

\begin{equation}
 \lbrack A^{+}(\lambda)A^{-}(\lambda),{\left(B^{+}\right)}^n(\lambda)\rbrack=
\sum^{n}_{i=1}R(\lambda+i){\left(B^{+}\right)}^n(\lambda).
\end{equation}
Hence the energy eigenvalues  of the Hamiltonian $\cal H_{-}$ are
given by

\begin{eqnarray}
\label{Enss} 
E^{(n)}_{-}&=&  \sum^{n}_{i=1}R(\lambda+i),
 \nonumber\\
&=&
\frac{-E^2\gamma^2(E)}{(\lambda+n)^2}+\frac{E^2\gamma^2(E)}{\lambda^2},
 \quad\gamma_n(E)=A_1+\frac{MA_2}{E_n}.
\end{eqnarray}
Thus, from equations (\ref{k}), (\ref{a}), (\ref{EA+A-}) and
(\ref{Enss}),
 the energy eigenvalues associated to the lower component
$\hat{F}^n$ are given by

\begin{equation}
E^{(n)}= \sqrt{\frac{M^2}{1+\frac{\gamma^2_n}{(\sqrt{k^2 -
\gamma^2_n}+n)^2}}}, \quad n= 0, 1, 2,\cdots.
\end{equation}
Solving the last equation we get

\begin{eqnarray}
{E_{\hat n}\over M}=-{A_1A_2\over ({\hat n}^2+
{A_1}^2)}\pm\left[{A_1^2A_2^2\over \left({\hat
n}^2+{A_1}^2\right)^2}+ {{\hat n }^2-A_2^2\over {\hat
n}^2+{A_1}^2}\right]^{1\over 2}
\end{eqnarray}
where $\hat n = n_r+\lambda $ and $n_r= 0,1,2, \cdots$. The
corresponding eigenfunctions associated to the lower component are
given by

\begin{equation}
\hat{F}_n(r)=(B^+)^n\hat{F}_0(r)
\end{equation}
where $\hat{F}_0$ is given in Eq. (\ref{F0}).

From the previous equation we can determine the first excited state of the eigenfunction. Using

\begin{equation}
B^+(\lambda) = T(\lambda)A^{+}(\lambda)
\end{equation}
and according to the equations (\ref{AA+-}), (\ref{B-}) and (\ref{T}) found

\begin{equation}
B^+(\lambda) = \left(\frac{d}{dr}+\frac{\lambda}{r} -\frac{EA_1 + MA_2}{\lambda}\right)e^{\frac{\partial}{\partial\lambda}}.
\end{equation}

Hence, for the first excited state we get the corresponding eigenfunction associated to the lower component 

\begin{equation}
F_1(r) = B^+\hat{F}_0(r) = \left(\frac{d}{dr}+\frac{\lambda}{r} -\frac{EA_1 + MA_2}{\lambda}\right)e^{\frac{\partial}{\partial\lambda}}
r^{\lambda}e^{-\frac{(E_0 A_1 + MA_2)r}{\lambda}},
\end{equation}
so,

\begin{equation}
F_1(r) = N_1\left(\frac{1}{r} - \frac{EA_1 + MA_2}{\lambda(\lambda + 1)}\right)r^{\lambda + 1}e^{-\frac{(EA_1 + MA_2)r}{\lambda + 1}},
\end{equation}
where $N_1$ is a normalization constant for the first excited state.


\section{Conclusion}

In the non-relativistic supersymmetric quantum
mechanics formalism, ladder operators are constructed to obtain the
complete set of the bound state energy spectrum and eigenfunctions for a
relativistic potential given by a Coulomb-like and
a Lorentz scalar potential
in the Dirac equation; one adopts polar coordinates. These new generalized ladder operators
obtained via supersymmetry shape-invariant Hamiltonians  
in quantum
mechanics can be reduced to the ones of the Coulomb potential, which provides
us the exact relativistic energy. These Hamiltonians, ${\cal H}_{-}$ and
${\cal H}_+$, are SUSY-partner Hamiltonians with identical energy spectra,
except for the ground state level, satisfying 
two Schr\"odinger-like
wave equations, with the vector and scalar potentials.

The vastly
simplified algebraic treatment within the framework of the primary SUSY
algebra, in terms of conserved operator (Runge-Lenz operator) in the full functional space for the nonrelativistic Coulomb problem with spin has been investigated \cite{TT93}. The primary SUSY with Johnson-Lippmann \cite{JL50} operator
for the Dirac-Coulomb problem is
not possible \cite{Stahlhofen97}. 

Our results on the connection between the generalized Johnson-Lippmann operator and SUSY for the bound eigenstates of the generalized relativistic Coulomb problem with position dependent mass will be reported separately.

\vspace{1cm}

\centerline{\bf Acknowledgments}

RLR wishes to thank the staff of the CBPF and UAE-CES-UFCG
for
the facilities and are also  grateful to  J. A. Helayel Neto 
for interesting discussions and suggestions. ESR and AFL would like to acknowledge  the Post-Graduate
Coordination in Physics of the  UAF-UFCG by incentives and encouragement.
RLR is also grateful for interesting discussions with Arvind Narayan Vaidya
(In memory), whose advises and encouragement were fundamental.


\end{document}